\documentclass[12pt]{article}
\usepackage{amssymb}
\usepackage{amsmath}
\usepackage{epsfig}
\newcommand{\be}{\begin{equation}}
\newcommand{\ee}{\end{equation}}
\newcommand{\bea}{\begin{eqnarray}}
\newcommand{\eea}{\end{eqnarray}}
\newcommand{\m}{\mathbf}
\date{}
\begin{document}

\begin{center}
{\large\bf Prospects for study of the effect of the electronic screening on the $\alpha$ decay in the storage rings}

\large
\bigskip
{F. F. Karpeshin$^1$, M. B. Trzhaskovskaya$^2$, and L. F. Vitushkin$^1$ }\\
      $^1$D. I. Mendeleyev Institute for Metrology, Saint-Petersburg, Russia

     $^2$National Research Center ``Kurchatov Institute'' --- Petersburg Nuclear Physics
      Institute, Russia

 \vspace{0.1cm}
{\it E-mail: fkarpeshin@gmail.com}
\\
\end{center}

\begin{abstract}Study of the role of the electron screening in the $\alpha$ decay is advanced, aimed at the experimental test in the storage rings. To this end, systematic calculation of the effect in heavy ions of the nuclei within the Ra to Po domain of the alpha emitters is conducted within the adiabatic approach. The obtained effect is to slow down the decay by an amount within the percent value. It is of the opposite sign comparing to what is predicted within the conventional frozen shell model. The reason for this divergence is demonstrated. Testing this difference experimentally in storage ring facilities is proposed. It is of great interest for advancing the astrophysical fundamental research.
\end{abstract}

\large

\section{Introduction}
      The $\alpha$ decay plays a significant role both in practical applications and scientific research, starting from astrophysics (e.g. \cite{salpeter}) and up to the laboratory research in plasma physics. At the same time, there is a contradiction between the laboratory study, which considers nuclear reactions with no respect to a possible role of the electron shell or environment, and  applications, which deal with various electronic shells or environments. Specifically, the $\alpha$ decay affects the nuclear synthesis in the stars. The extent of the screening effect in ultra-high-density stellar environments might become
significant and would deserve further investigation. The beginning was laid, for example, in Refs. \cite{itl}.

       Another astrophysical issue is related with the time reversibility of the nuclear reactions occurring in the interior of stars,  whose study is one of the most promising areas of modern nuclear physics. Most of these reactions encounter with the similar problem, as  the $\alpha$ decay, aggravated by  extremely small cross-sections at small energies, being still not available for direct measurements in the laboratory. Consequently, indirect approaches are developed in order to better know their
      cross-sections and rates,
      such as the asymptotic normalization coefficients, or the Trojan horse method  (e.g. \cite{bloh}). From this viewpoint, direct information on the rates and role of the electron screening in the reverse process of $\alpha$ decay becomes important.

      Surprisingly, up to paper \cite{prc} of 2013, everybody considered this question within the framework of the frozen electron shell (FS) model (e.g. \cite{erma,perl,PR2,Zin,patyk} and Refs. cited therein), although the electrons are four orders of magnitude lighter than the $\alpha$ particles, and are certainly strongly affected by the $\alpha$ particle slowly traversing the shell. The calculated results were even of different signs, in spite of simple arguments, which can be readily told out based on the physical ground \cite{yaf}. They suggest that the electron environment retards the decay. This might promote formation of heavy elements and actinides in the $r$ process in the stars. The problem is therefore to involve the interaction of the $\alpha$ particle with the electronic shell into the consideration. The main difficulty, which however arises on this way, is in  treating the interplay of the two different scales involved. From one side, one has to take into account a decrease of
$\Delta Q \approx$ 40 keV in the energy of the $\alpha$ particle. This arises and is being finally formed on the distances of the atomic scale. It is expected to act as to suppress the decay (\cite{erma}). In contrast, from the other side, $\alpha$ decay is predetermined by the strong short-range $\alpha$-nucleus interaction. Its influence is ruled out on the outer turning point of the Coulomb barrier, which is essentially inside the electron shell, tens fermi from the nuclear surface. As a result, an intuitive expectation is that the shell should not affect the probability significantly. Moreover, in this region the electrons produce a negative potential for the $\alpha$ particle, which may be assumed to facilitate the decay. Consecutive account of all these factors can be only performed within the adiabatic approach, which was proposed in Ref. \cite{prc}. Analogy was noted with suppression of prompt fission in muonic atoms of actinides. That example teaches that it is not the electronic potential  so important but its gradient, determining the force acting on the $\alpha$ particle.  Important for experimental research details were elaborated in Refs. \cite{yaf,therm}. Furthermore, the adiabatic approach was recognized and  extended in Refs. \cite{dzuba,chin}. Herein, we advance the study of the effect of the electron screening and discuss a feasibility of experimental test for it, using such contemporary facilities like storage rings, available at GSI and IMP Lanzhou. In the next section, we remind the principles of the model. Numerical results for the isotopes of Po to Ra domain are obtained in Section \ref{num}. Prospects of the experimental study are discussed in Section \ref{exp}. Resume is derived in the concluding section \ref{resume}.

\section{Remind of the model: physical premises}

      Within the framework of the Gamow theory, a conventional expression for the $\alpha$ decay probability is essentially given by the product of two factors: the cluster preformation and the penetration probabilities. The former is assumed  not be affected by the shell. The latter factor $P$ is determined by action $S$ as follows:
\be
P=e^{-2S} \,,
\ee
 with
\be
S = \int_{R_1}^{R_2} \sqrt{2m[E-V(R)]}dR \,. \label{BN}
\ee
Here $V(R)$ is the potential energy of interaction of the emitted $\alpha$ particle with the rest of the system, including interaction with the nucleus and the electronic shell in the case of atomic system. $m$ is the mass of the $\alpha$ particle, and $E$ is its kinetic energy in the asymptotic region where $V(R)$ can be neglected. Furthermore,
$R_1$, $R_2$ are the turning points. The effect of the electron screening is then expressed as
\be
Y = P_a/P_n-1\,,   \label{effect}
\ee
where subscripts `n' and `a'  indicate either the case of bare nuclei or atoms, respectively.
The problem of consistent taking into account the aforementioned factors became a stumbling block for many early calculations. The study of Patyk {\it et al.} \cite{patyk}, who performed the most consistent calculations of alpha decay for the chain of radon isotopes, put a period to the development of the FS model.

\begin{figure}
\includegraphics[width=0.8\textwidth]{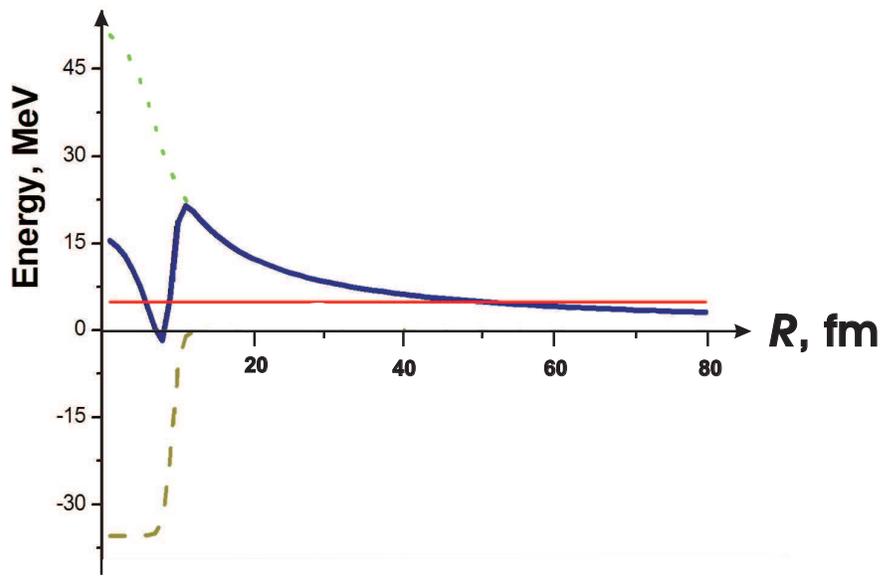}
\caption{\footnotesize Potential energy of the $\alpha$ particle interaction with the nucleus (thick full curve), resulting from the superposition of the strong attractive potential well (dashed curve) and the repulsive Coulomb interaction (dotted line).}
\label{nucpf}
\end{figure}

In more detail, in the case of  bare nuclei,  the potential energy consists of the strong short-range interaction $V_n(R)$, centrifugal repulsion $V_{cf}(R)$ and Coulomb attraction $V_{Coul}(R)$. It reads as follows:
\be
V(R) = V_n(R)+V_{cf}(R)+V_{Coul}(R) \equiv V_N(R)\,.  \label{BNpot}
\ee
Nuclear potential $V_N(R)$ calculated for $^{226}$Ra, with the parameters from \cite{denis}, is illustrated in Fig. \ref{nucpf}. It looks as a shallow well, formed by the superposition of the Coulomb repulsion and the strong potential well $V_n(R)$ in the nuclear vicinity, and a flat wide slope formed by $V_{Coul}(R)$ at more distant $R$.  The straight line  shows the energy of the emitted $\alpha$ particle. The centrifugal potential $V_{cf}(R)$ arises in the case of the angular momentum of the $\alpha$ particle $L\neq 0$, and is not shown in the Figure. It only slightly modifies the shape of the barrier, somewhat changing the effect of the electron shell in the third decimal.

      In atoms, the adiabatic potential energy of the $\alpha$ interaction with the shell
$U_{e-\alpha}^{\text{ad}} (r)$  is added:
\be
V(R) \equiv V_a(R) = V_N(R)+ U_{e-\alpha}^{\text{ad}} (R)  \,.
\ee
Furthermore, the $Q$ value is diminished by $\Delta Q$ because of the  rearrangement of the shell in the daughter atoms, with the related change of the total binding energy $\Delta B\equiv \Delta Q$.
The resulting expression for the action integral becomes as follows:
\be
S_a = \int_{R_1'}^{R_2'} \sqrt{2m[E-V_N(R) + \Delta Q -  
U_{e-\alpha}^{\text{ad}} (R)]} dR  \,. \label{Ap}
\ee

      In order to better realize the difference between the adiabatic and FS models, let us dwell in more detail, than it is done in Refs. \cite{prc,yaf}, on the construction of     the 
$e-\alpha$ prompt electrostatic potential of interaction of the $\alpha$ particle with the electronic shell 
$U_{e-\alpha}^\text{ad}(R)$. It is one of the main constituents of the adiabatic method. Let us put down a general expression as follows: 
\be
U_{e-\alpha}^\text{ad} (R) = -\zeta e^2 \int \frac{\rho_e(r)}{|\m r-\m R|}\ d^3r + \text{const}\, \label{Uad}
\ee  
with $\rho_e(r)$ being the prompt electron density,  and 
$\zeta$ = 2 --- charge of the $\alpha$ particle. $\rho_e(r)$  depends on $R$. It can be considered  as spherically-symmetric at $R<R_s$, $R_s$ being the point of appearance of the  $\alpha$ particle on the nuclear surface. Its value is usually calculated by means of the self-consistent method (e.g. \cite{RAIdens}):
\be 
\rho_e(r)=4\pi\sum_i N_i [G_i^2(r)+F_i^2(r)] \,,  \label{dens}
\ee
where $G_i(r)$ and $F_i(r)$ are the radial Dirac electron wave functions, normalized as follows:
\be
\int_0^\infty [G_i^2(r)+F_i^2(r)]dr =1\,. 
\ee
 Summation in (\ref{dens})  is performed over the shells $i$, with $N_i$ being the occupation numbers. We introduced  an arbitrary constant in Eq. (\ref{Uad}), which we define later. By definition, $e-\alpha$ interaction potential in the FS model is obtained with const = 0. 
Starting from this moment, at $R>R_s$,  $G_i(r)$ and $F_i(r)$  depend on $R$. This is an important element constituting difference between the FS model and the adiabatic approximation. 
Increasing  $R$ also increases the r.m.s. radius of the total nuclear charge, which causes diminishing the binding energy of the atom and rise of the electronic term \cite{prc,yaf}. Furthermore, the electronic density gradually becomes   aspherical, following  the motion of the $\alpha$ particle. For the calculation, in the first order of the perturbation theory this effect can be neglected at small $R$ around the nuclear surface, including a physically interesting subbarrier region. Then Eq. (\ref{Uad}) can be expressed as follows:
\bea
U_{e-\alpha}^\text{ad} (R) = 4\pi \phi(R) + \text{const}\,, \label{U}\\
\phi(R)=  u(R) - \frac{\zeta ze^2}{R_e} \,, \nonumber\\  
u(R) =  -\zeta e^2\sum_i N_i \int_0^R[G_i(r)^2+F_i^2(r)](\frac 1R-\frac1r)dr\,,  \label{Ni}  \\
\frac {1}{R_e} = \sum_i N_i \int_0^\infty [G_i^2(r)+F_i^2(r)]\frac{dr}{r} \,.   \label{Re}
\eea
Here we denoted $z$ the number of electrons in the atoms, not to be mixed with the atomic number $Z$. By definition, function $\phi(R)$ determines the potential in the FS model.
We normalize the constant by the natural boundary condition:
\be
U_{e-\alpha}^\text{ad} (R)\to 0 \quad \text{at} \quad R\to\infty\,.    \label{inf}
\ee
One cannot apply condition (\ref{inf}) to the expression (\ref{U}) yet for expansion (\ref{Ni}) is not valid at large $R$.

\begin{figure}
\includegraphics[width=0.8\textwidth]{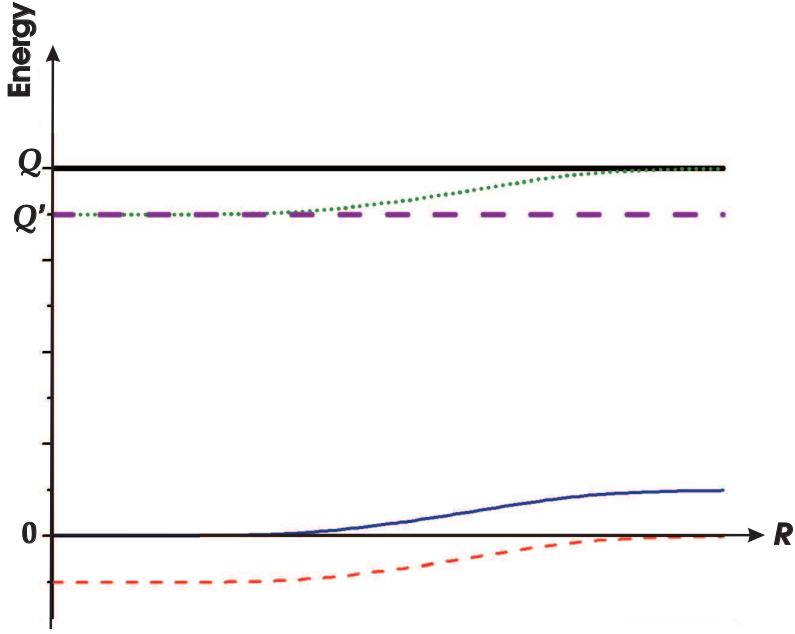}
\caption{\footnotesize Thick full curve shows the $Q$ value in the case of a bare nucleus;
dotted line --- the adiabatic $e-\alpha$ potential, which should be added in the case of a neutral atom, after which the alpha energy becomes $Q^\prime$. The variation of the adiabatic electronic potential equals the energy of the electronic rearrangement $\Delta B$. Dotted line shows that the difference $\Delta Q=Q-Q^\prime$ exactly equals $|\Delta B|$. The action in the case of neutral atoms does not change if the $Q^\prime$ value is shifted simultaneously with the electronic potential by the same value. The figure thus illustrates that the action for the atom can be considered as that for the bare nucleus, with the same $Q$ value, but adding the effective shifted electronic potential (thin solid line).}
\label{shiftf}
\end{figure}

      Note that at  the starting  point, the potential energy of $e-\alpha$ interaction is 
$U_{e-\alpha}^{\text{ad}} (R_s)$. Then, according to the Feynman-Hellman theorem \cite{FH}, within the framework of the adiabatic approach, the work $w(R)$ done by the $\alpha$ particles over the electron shell when moving from the starting point $R_s$ to a point $R$ acquires the usual expression depending on the potential difference at the end and start points:
\be
w(R) = \int_{R_s}^R \frac {d U_{e-\alpha}^{\text{ad}} (R') }{dR'}\ dR' =
U_{e-\alpha}^{\text{ad}} (R)-U_{e-\alpha}^{\text{ad}} (R_s)  \,.    \label{AadR}
\ee
The constant which is not defined yet cancels out in Eq. (\ref{AadR}). This equation with the account of  (\ref{inf}) allows one to express the  constant in terms of  $\Delta Q$. Going to the limit $R\to\infty$, one arrives at 
\be
\Delta Q = U_{e-\alpha}^{\text{ad}} (R_s) \,.    \label{Aad}
\ee
Now condition (\ref{Aad}) defines the constant:
\be
\text{const} = \Delta Q-\phi(R_s) \,.  \label{cn}
\ee
Substituting Eq. (\ref{cn}) into (\ref{Ap}), we see that the effect of the electron shell is reduced to the mere addition of a purely electronic effective potential \cite{prc}
\bea
W_{eff}^{\text{ad}}(R)=  4\pi[\phi(R)-\phi(R_s)] = 
\nonumber  \\
-4\pi\zeta e^2\sum_i N_i \int_{R_s}^R[G_i(r)^2+F_i^2(r)](\frac 1R-\frac1r)dr\,.  \label{eff}
\eea
Eq. (\ref{eff}) differs from Eq. (\ref{Ni}) by the low bound of integration. This result is illustrated in Figure \ref{shiftf}.
Involving  the effective electronic potential $W_{eff}^{\text{ad}}(R)$ resolves the conflict of scales mentioned in the Introduction. It appears as a perturbative positive definite extra potential, reasonably weak enough, which slightly affects the barrier penetration probability. Below, we will mean potential (\ref{eff}) under the alpha-electron interaction. The energy $E$ in formula (\ref{eff}) remains the same as in Eq. (\ref{BN}) for  bare nuclei. 

       The Feynman---Hellman theorem is not relevant in the FS model,  in view of that it is only valid for stable systems \cite{FH}, 
which obviously do not include a FS model. In this case, one can calculate the $e-\alpha$ interaction potential directly by means of Eqs. (\ref{U}) to (\ref{Re}) with const = 0, whilst holding the $\Delta Q$ value in Eq. (\ref{Ap}). For the purpose of comparison of the models, let us introduce the effective potential for the FS model $W_{eff}^\text{FS}(R)$ by mere inclusion of $\Delta Q$ value into $\phi (R)$:
\be
W_{eff}^\text{FS}= 4\pi\phi(R) -\Delta Q \,.   \label{effFS}
\ee
Then in the both models, the expression for the action integral reads as follows:
\be
S_a = \int_{R_1'}^{R_2'} \sqrt{2m[E-V_n(R)-V_{Coul}(R)- W_{eff}(R) } dR \,,  \label{AFS}
\ee 
with $W_{eff}(R)$ given by either (\ref{eff}) or (\ref{effFS}).
       
\begin{figure}
\centerline{ \epsfxsize=8cm \epsfysize=7cm
\epsfbox{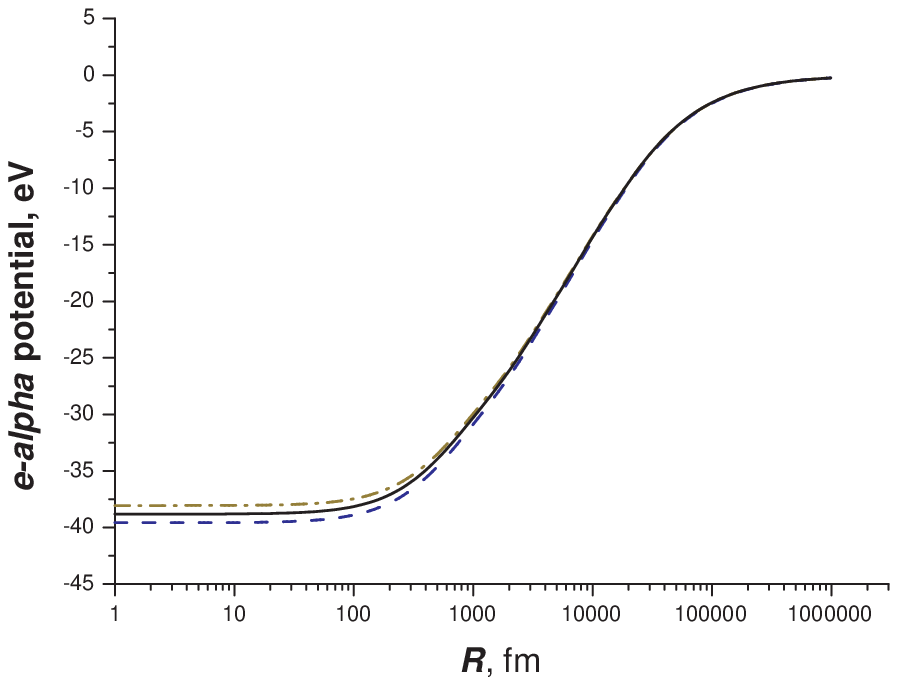} \epsfxsize=8cm \epsfysize=7cm
\epsfbox{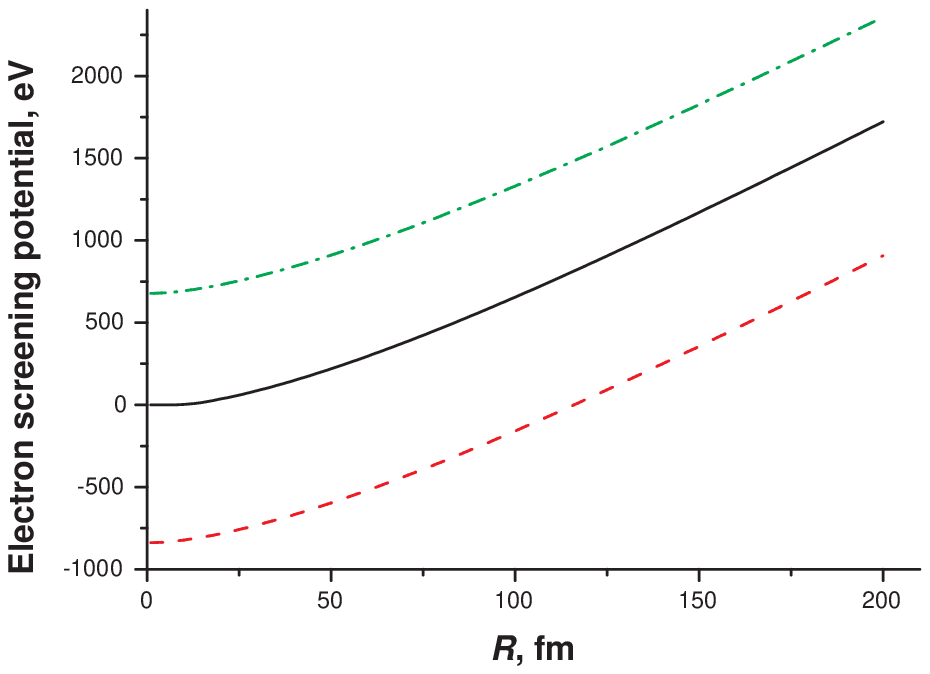} }
\caption{\footnotesize  {\it a} (left) --- schematic  $e-\alpha$ potentials in the case of decay of neutral $^{226}$Ra atoms: the   adiabatic approach (full curve), in comparison with the potentials of the frozen shell in the initial  (dashed curve) and final (dash-dotted curve) atoms, respectively; {\it b} (right) --- the effective calculated $e-\alpha$ potentials in the subbarrier region.}
\label{AdiabF}
\end{figure}

      In Figs. \ref{AdiabF} the resulting interaction potential of  the $\alpha$ particle with the electron shell is presented in more detail in the case of neutral $^{226}_{88}$Ra atoms $\alpha$ decay. The calculations are performed by means of Eqs. (\ref{U}) to (\ref{Re}), using the RAINE package of the computer codes \cite{RAINE}. Fig. \ref{AdiabF}{\it a} shows general view of the calculated potentials. 
At small $R$, the adiabatic potential   starts from 
$U_{e-\alpha}^{\text{ad}} (R_s) = \Delta Q$ at $R=R_s$. The $\Delta Q$ value is calculated as the difference of the total binding energies of the mother and daughter atoms. At large $R$, qualitative view of the curve is shown.  It vanishes at large $R$ as 
$-\zeta(z-\zeta)e^2/R$ in the main term. At small distances, calculation shows that the adiabatic potential turns out to be confined between the FS potentials of the atoms $Z$ and $Z-2$.  This is reasonable in view of that the real electronic density evolves from the parent to the daughter atoms. 

       Fig. \ref{AdiabF}{\it b} shows the effective potentials   in the subbarrier region.  The adiabatic potential starts from  zero value. The  effective FS potentials start from the negative and positive values in the initial and final atoms, respectively.
Numerically, in the case of $^{226}$Ra, 
$U_{e-\alpha}^\text{FS}(0) $  = $-$39.572 keV was
obtained, and $\Delta B$= $-$38.805 keV. Thus,  the FS curve starts from $-$0.767 keV at $R$ = 0. This is in a close agreement with Ref. \cite{patyk}, where the value of about 750 eV can be concluded (cf. Fig 1 of Ref. \cite{patyk}). According to Ref. \cite{Rodrigues},  $\Delta B$     = 640927 $-$ 679616 = $-$38789 eV, that is by only 16 eV less than our value.  The left and right barrier turning points are $R_1$ = 9 fm, $R_2$ = 51 fm. The three conclusions drawn below clearly follow  fig. \ref{AdiabF}:

1) The effect of the shell on the $\alpha$ decay rate is negative (as the adiabatic effective potential is represented by a  positive definite curve).

2) Contrary, within the framework of the FS model, the effect is positive (as the corresponding lowest curve really lies below zero in the subbarrier area).

3) In the subbarrier region, the absolute value of the adiabatic potential is considerably less than that of the FS model.  This explains why the effect within the consecutive adiabatic approach turns out not only to be of the opposite sign, but also considerably less in absolute value.

\begin{figure}[!hbt]
\centerline{ \epsfxsize=8cm \epsfysize=8cm \epsfbox{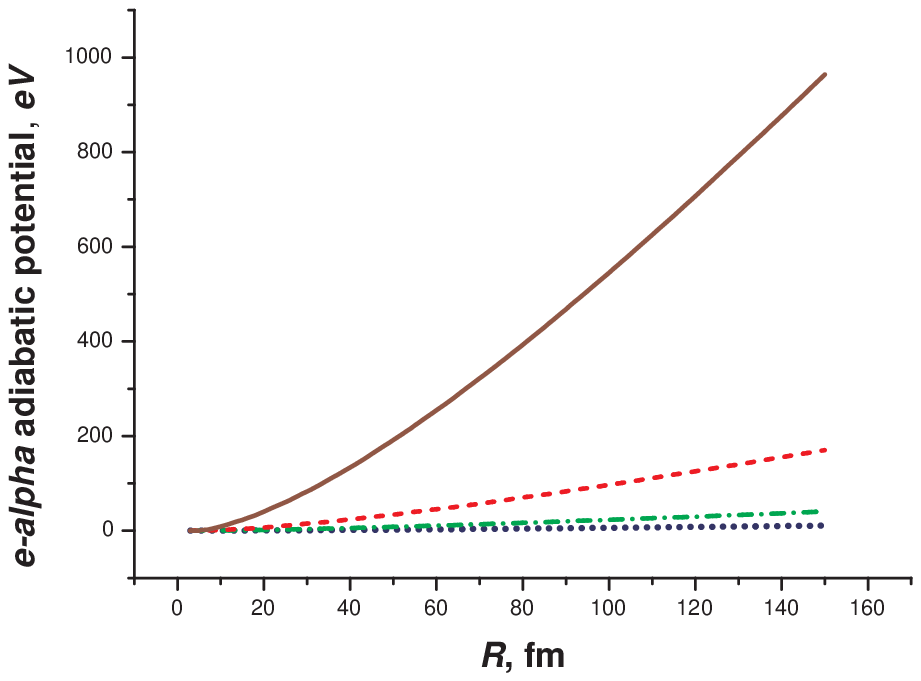}
\epsfxsize=8cm \epsfysize=8cm \epsfbox{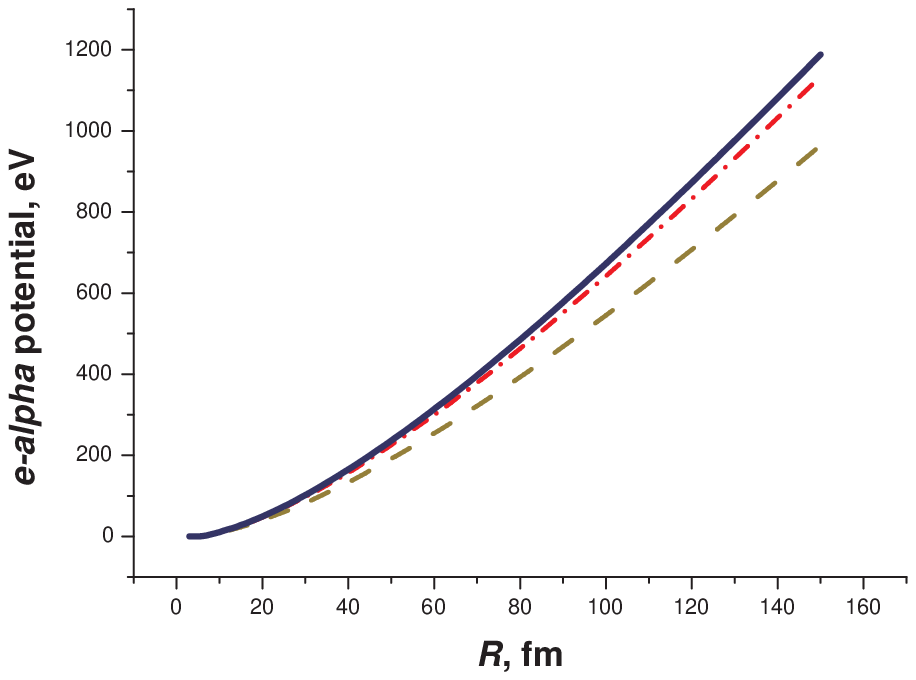} } \caption{\footnotesize Contributions of separate shells into the
$e-\alpha$ adiabatic potential in the case of neutral Ra atoms. Left: $K$
shell --- full curve, $L$ shell --- dotted curve, $M$ shell ---
dash-dotted curve, and $N$ shell --- dotted curve. Right: the
$e-\alpha$ adiabatic potential of the $He$-like ions (dashed
curve), Ne-like ions (dash-dotted curve), and neutral atoms (full
curve). }
\label{seprtf}
\end{figure}

    Moreover,  the adiabatic approach allows one to consistently calculate the contribution of each electron shell to the overall effect. This is performed directly by means of Eq. (\ref{eff}). Results are presented in Fig. \ref{seprtf}. It follows from Fig. \ref{seprtf} that about 80\% of the total screening potential is expected to the contribution from the $K$-shell electrons. This gives an opportunity of performing an accurate and beautiful experiment, which we will discuss in the next section.

\section{Results}
\label{num}

    Representative calculations were performed in Refs. \cite{prc,yaf,therm} for various $\alpha$ emitters throughout the periodic table, and with different decay energies. They show that the effect of screening strongly decreases with increasing the $Q$ value and decreasing the lifetime. Within the adiabatic approximation, it is evident that the  inner electrons produce more effect, as they are more sensitive to the motion of the $\alpha$ particle in the subbarrier area near the nucleus. According to Fig. (\ref{seprtf}), more than 80\% of the effect are produced by the $K$ electrons. This suggests an elegant and basic way of experimental check {\it e.g.} through measurement of the difference in the decay rate between the He-like and bare ions of the same nuclei --- alpha emitters. Monochromaticity parameters of the storage ring beam are good enough, in order to detect the recoil nuclei  by the Schottky method \cite{friz}.

        Consider dependence of the decay rate on stripping the electron shell in more detail. Results of calculation for $^{226}$Ra and $^{212}$Rn atoms are presented in Table \ref{elt}.
\begin{table}
\caption{\footnotesize   The effect of the electron screening $Y$ as calculated for different electronic configurations of the $^{226}$Ra atom}
\begin{center}
\begin{tabular}{||c|c||c|c|c|c|c||}
\hline   \hline
Nuclide & $Q$, MeV & Neutral & $K$-shell & $K$- and $L$-shells & He-like ion & Ne-like ion \\
\hline
$^{226}$Ra & 4.87063 & $-$0.246 & $-$0.200 &   $-$0.235  & $-$0.201 &  $-$0.236   \\
$^{212}$Rn & 6.385 & $-$0.086 & $-$0.070 &  $-$0.081  &  $-$0.071 &  $-$0.082   \\
\hline \hline
\end{tabular}
\end{center}
\label{elt}
\end{table}
The $Q$ values (column 2) are cited according to Ref. \cite{lederer}. In  column 3, the $Y$ values (\ref{effect}), calculated within the adiabatic approach, are presented for the neutral atoms. The partial contributions from the $K$ shells and both the $K$ and $L$ shells, respectively, are listed in columns 4 and 5. In the sixth and seventh columns the results are listed for the He-like and Ne-like ions, respectively. It follows from these results that 82\% of the effect are due to the contribution of the $K$-shell electrons. 96\% --- nearly full effect --- is achieved in the case of
Ne-like atoms. Note that in the case of He-like or Ne-like ions the effect is a little greater than if calculated for neutral atoms with allowance for only the $K$- or both $K$- and $L$-shells, respectively. This fact has a simple explanation on the physical ground. In the ions, the electronic orbitals are more compact around the nuclei, and therefore the wavefunctions are greater in the area under the Coulomb barrier. This consequently causes bigger values of the integrand in Eq. (\ref{eff}) and the related increase of the  effect.

      The experiment can be realized in a similar way to which was applied in search for the time modulation in beta decay of Pm ions \cite{Pm}. The daughter product from the $\alpha$ decay will stay in the ring and should be seen by Schottky analysis. If one starts from $^{212}$Rn$^{84+}$, one should have most of the time $^{208}$Po$^{82+}$, it will be good to detect both. If one uses bunches of ions then the Schottky needs to be calibrated and one has to find out how well this can be done for the four cases of decay from $^{212}$Rn$^{84+}$ and $^{212}$Rn$^{86+}$. Concerning the ``background" arising by ionization and shake-off from Rn$^{84+}$, and there could also be electron capture by Rn$^{86+}$, these could be made quite weak.

\begin{figure}
\includegraphics[width=\textwidth]{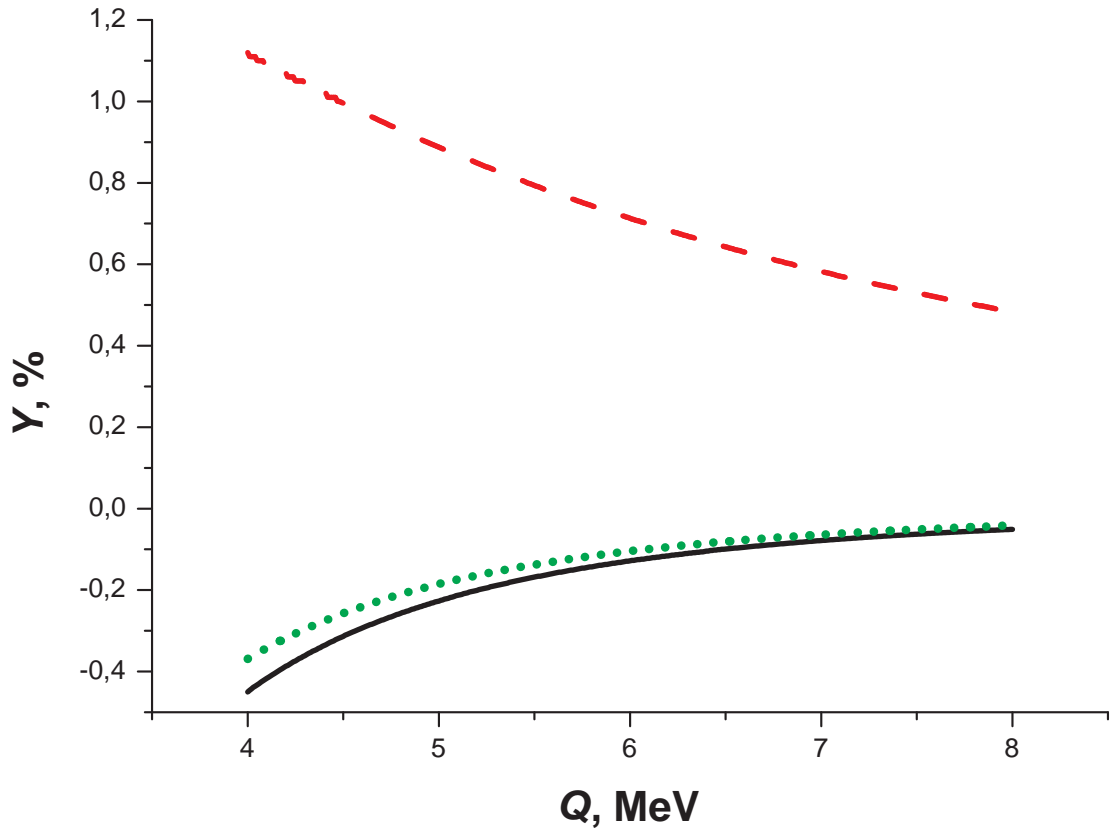}
\caption{\footnotesize Calculated relative change of the $\alpha$ decay rate $Y$ in neutral atoms as compared to that of bare nuclei in the isotopes of Ra (full line) and Rn (dotted line) $Y$ against the released energy $Q$. Dashed curve --- the same for Ra isotopes as calculated in the FS model.}
\label{Yf}
\end{figure}

      On one hand, the $\alpha$ lifetimes of seconds to minutes seem  feasible from the viewpoint of experiment using storage rings. On the other hand, the calculated values of $Y$ are mainly determined by the $Q$ values. They weakly depend on the atomic and mass numbers $Z$, $A$ of the nuclides within the
Ra -- Po domain. In Fig. \ref{Yf} we present the $Y$ values as calculated for the isotopes of Ra and Rn against their $Q$ values in neutral atoms. For comparison, the results, obtained within the FS model, are also shown in the same figure.

      In order to have a big effect, it is desirable to use the isotopes with small $Q_\alpha$. But   small $Q_\alpha$ correlate with long halflives which may comprise days and years, which is inconvenient for measuring in the storage rings. In Table \ref{Rat} we  present the resulting $Y$ values  for radium isotopes with halflives within seconds to minutes and the $\alpha$ branching ratios 100\% or close. Most abundant isotope $^{226}$Ra and the next $^{222}$Ra are characterized with a feasible effect $Y$ = $-$0.25 and $-$0.14 percent, respectively, with comparatively long halflives. Among the lighter isotopes,
there are two groups around $A\approx$  222 and 213. They are characterized with the  $Q_\alpha$ values of 6--7 MeV, and the effect $|Y|\lesssim$ 0.1\%. Results for other such suitable isotopes of Fr, Rn, Ac, At and Po are listed in Table \ref{Yt}.
      \begin{table}
      \caption{\footnotesize Relative change of the decay rate $Y$ between the decay in bare nuclei and He-like ions  in the case of Ra isotopes }
      \begin{center}
      \begin{tabular}{||c||c|c|c||}
      \hline   \hline
       $A$ & $Q_\alpha$ & $T_{1/2}$ & $Y$, \% \\
      \hline
      226 & 4.87063 & 1600 y & $-$0.246  \\
      224 & 5.78887 & 3.66 d & $-$0.14 \\
      222 & 6.681 & 38.0 s & $-$0.091  \\
      221 & 6.884 & 28 s & $-$0.083 \\
      220 & 7.595& 18 ms & $-$0.060\\
      214 & 7.273 & 2.46 s & $-$0.069 \\
      213 & 6.861 & 2.74 m &  $-$0.084  \\
      212 & 7.0319 & 13.0 s  &  $-$0.071 \\
      \hline   \hline
      \end{tabular}
      \end{center}
      \label{Rat}
      \end{table}
      \begin{table}
      \caption{\footnotesize Relative change of the decay rate $Y$ as in Table \ref{Rat}  calculated for  other elements }
      \begin{center}
      \begin{tabular}{||c||c|c|c||}
      \hline    \hline
       Nuclide & $Q_\alpha$ & $T_{1/2}$ & $Y$, \% \\
      \hline
      $^{221}$Fr & 6.4579 & 4.9 m & $-$0.10 \\
      $^{220}$Fr & 6.8007 & 27.4 s & $-$0.086 \\
      $^{213}$Fr & 6.9051 & 34.6 s & $-$0.082 \\
      $^{220}$Rn & 6.40467 & 55.6 s & $-$0.10 \\
      $^{219}$Rn & 6.9461 & 3.96 s & $-$0.080  \\
      $^{212}$Rn & 6.385 & 23.9 m & $-$0.11  \\
      $^{223}$Ac & 6.7831 & 2.10 m & $-$0.087 \\
      $^{222}$Ac & 7.1374 & 5 s & $-$0.074  \\
      $^{202}$At & 6.3537 & 184 s & $-$0.11 \\
      $^{201}$At & 6.4733 & 89 s & $-$0.10 \\
      $^{200}$At & 6.5964 & 43 s & $-$0.095 \\
      $^{199}$At & 6.780 & 7.2 s & $-$0.087  \\
      $^{198}$At & 6.893& 4.2 s & $-$0.083 \\
      $^{197}$At & 7100 & 0.35 s & $-$0.075  \\
      $^{196}$At & 7200 & 0.3 s & $-$0.072  \\
      $^{211}$Po & 7.5945 & 0.516 s & $-$0.061 \\
      \hline   \hline
      \end{tabular}
      \end{center}
      \label{Yt}
      \end{table}

\section{Discussion}
\label{exp}

      Although the predicted magnitude of the effect is as small as about $10^{-3}$, there is a circumstance favorable for experimental studies: about 80\% of the effect is due to the contribution of the $K$-shell electrons. In order to observe the effect, it is therefore sufficient to compare the decay halflives for bare nuclei with the halflives for respective one- and (or) two-electron atoms, without involving neutral atoms --- their accumulation in accelerator rings is impossible. And generally, it is more reliable to observe small effects via difference measurements at the same facility. Thus, comparative measurements can be performed in the same storage rings with bare nuclei and respective ions of different multiplicity --- for example, helium-like, or neon-like ions.
    An experiment on comparison of the alpha decay half-lives in neutral atoms and their ions was already tested at GSI a few years ago \cite{noci}. H-like $^{213}$Fr ions were produced and injected in the storage ring.
The decreasing number of parent ions has been observed by the Schottky technics after each injection. However, because of a large statistical uncertainty, the measured half-life, $T_{1/2}$ = 34(6) s at rest, was compatible with the neutral half-life.
Analyzing that experiment  in the light of the present consideration, the following conclusions are worthy of noting.

      First, the main idea of that experiment sounds striking: comparison of the lifetimes of neutral atoms with those in H-like ions. Why not just bare nuclei? One $1s$ electron already produces close to half the effect.

    Second, they compared the lifetimes in the H-like ions with those in neutral atoms, as measured by traditional techniques in stoppers. In the case of $^{213}$Fr two independent neutral
half-life measurements have been carried out, one at GSI
and the other at the LNS-INFN. Even though they agree,
a discrepancy has been found with the value present in the
literature. Our present proposal excludes the development of such an unforeseen situation. It assumes that both lifetimes --- of the bare nuclei and He-like ions --- will be measured in the same channel of the storage ring by the Schottky method. This way will considerably compensate systematic uncertainties, which will be merely the same, whereas in the previous experiment they could be of different origin and therefore, not correlated with the measurements of the ionic half-life.

    Third, a general idea was told out in Ref. \cite{noci}  that such an experiment is feasible with the resulting uncertainty within 0.1\%, which is a typical value of the effect calculated herein. Therefore, one can think this estimation opens green light to realization of the experiment as discussed  above. Moreover, one can speak about experimental test  for correctness of either the FS model or adiabatic approach. This is even  more feasible, as one should choose between the estimates of 10$^{-2}$ and $-10^{-3}$, resulting from the FS and adiabatic models, respectively.

\section{Conclusion}
\label{resume}
      We consecutively considered the effect of the electron screening on the $\alpha$ decay rate from the viewpoint of its experimental investigation. The results presented previously demonstrate that within the adiabatic approach  the effect of the electron screening on the $\alpha$ decay rate is certainly negative, in spite of the attractive potential provided by the shell. The physics arguments in favor of this conclusion can be formulated as follows.

1) Raising electron terms create resisting force acting from the shell on the $\alpha$ particle.

2) Experimentally observed drastic suppression of prompt fission in muonic atoms of actinide elements.\\
 The main error of the FS model is in the identification  of the inner electrostatic Coulomb potential times 2$e$ with the potential energy of the $\alpha$ particle in its motion through the electronic shell \cite{itl,patyk}.  Non-identity  of these concepts is demonstrated in the best way by addition of the constant into Eq. (\ref{Uad}). Quint essence of the adiabatic approach is expressed by Eq. (\ref{Aad}). This condition results in the negative effect in the adiabatic approach and shows the reason for the positive effect within the FS model.  Flight of $\alpha$ particle through the electron shell rearranges the latter. In turn, this weakens the  effect of the shell  on the particle. Using the classic analogy, one can say that if the electrons are not fixed firmly on their places, then they take part of the momentum of the $\alpha$ particle and are partially carried  by the particle. It is  figuratively said in Ref. \cite{yaf} that the alpha particle is accompanied with a tail of the atomic electrons when it crosses the shell. Such a loosed dynamics leads to softening of the electric field strength around the $\alpha$ particle in comparison with what is expected in the FS model.

       General effect of retardation of the decay provided by the adiabatic approach is at the level of 10$^{-3}$.  FS model results in the effect of the opposite sign and percent value.  Its detection could be a difficult task. However, contemporary technique makes the observation of the effect, and all the more testing  for the difference between the models quite feasible. Moreover, application of the storage ring facilities makes unnecessary use of any detectors. The Schottky analysis allows one to conduct such an experiment. It needs no counters or other special technics apart from the beams of the target nuclei or ions. Minutes or even seconds of the beam time is enough for one experimental run. Such an impressive progress is provided by development of the experimental capabilities.  Inferring the experimental results will certainly put a milestone on the way of astrophysical research of the processes occurring in the stellar plasma, including the Sun.

      \section*{Acknowledgements} The authors are grateful to  X. Ma, M. Steck  and R. Schuch for many fruitful discussions of the questions considered herein. They want to thank T. St{\"o}lker and Yu. Litvinov for helpful remarks.

\end{document}